# Local Flaw Detection with Adaptive Pyramid Image Fusion Across Spatial Sampling Resolution for SWRs

Siyu You, *Student Member, IEEE*, Huayi Gou, Leilei Yang, *Graduate Student Member, IEEE*, Zhiliang Liu, Senior *Member, IEEE, and* Mingjian Zuo, *Fellow, IEEE*

*Abstract*—The inspection of local flaws (LFs) in Steel Wire Ropes (SWRs) is crucial for ensuring safety and reliability in various industries. Magnetic Flux Leakage (MFL) imaging is commonly used for non-destructive testing, but its effectiveness is often hindered by the combined effects of inspection speed and sampling rate. To address this issue, the impacts of inspection speed and sampling rate on image quality are studied, as variations in these factors can cause stripe noise, axial compression of defect features, and increased interference, complicating accurate detection. We define the relationship between inspection speed and sampling rate as spatial sampling resolution (SSR) and propose an adaptive SSR target-feature-oriented (AS-TFO) method. This method incorporates adaptive adjustment and pyramid image fusion techniques to enhance defect detection under different SSR scenarios. Experimental results show that under high SSR scenarios, the method achieves a precision of 94.73% and a recall of 96.77%. It remains robust under low SSR scenarios with a precision of 94.30% and recall of 97.32%. The overall results show that the proposed method outperforms conventional approaches, achieving state-of-the-art performance. This improvement in detection accuracy and robustness is particularly valuable for handling complex inspection conditions, where inspection speed and sampling rate can vary significantly, making detection more robust and reliable in industrial settings.

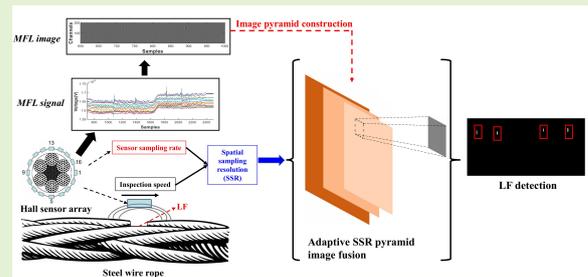

## I. Introduction

STEEL wire ropes (SWRs) are critical components in many industries like lifting, transportation, mining, and construction, where they are used for carrying heavy loads and enduring extreme mechanical stress [1], [2], [3], [4]. Due to their vital role in ensuring operational safety and efficiency, the regular inspection of SWRs is essential to prevent failures that could lead to hazardous situations or equipment damage [5]. SWRs are subject to various types of defects over time, mainly due to mechanical wear, environmental factors, and fatigue [3]. Common defects in SWR include wire breaks, corrosion, and fatigue cracks, all of which are types of local flaws (LFs) [1], [6]. LFs refer to localized damage or small cracks within the SWR, typically caused by factors such as fatigue or corrosion [7]. Unlike more widespread defects, LFs are often subtle and challenging to detect, yet they pose a disproportionate risk to the structural integrity of SWR. LFs can significantly impair the load-bearing capacity and longevity of the SWR, and if left undetected, may lead to catastrophic failures, posing severe safety risks in applications that rely on SWR.

Given the importance of identifying LFs early, non-destructive testing (NDT) methods have become a crucial tool for the inspection and maintenance of SWR. NDT techniques enable defect detection in materials and structures without causing damage [8]. Among these, magnetic flux leakage (MFL) stands out for its efficiency and stability. The principle behind MFL is that when a SWR is magnetized, the LF disrupts the magnetic field [9]. These disruptions result in magnetic flux leakage, which can be detected using sensors such as Hall sensors. Based on existing research, LF detection methods can be generally divided into 2 categories: signal-based method and image-based method.

Signal-based methods refer to those methods that diagnose LFs based on the characteristics of MFL signals. For instance, Kim et al. focus on analyzing the morphological features of the MFL signal, such as the peak-to-peak value, the peak value of

Manuscript received x Month 20xx. This work is supported by the Sichuan Science and Technology Program under Grant 2024JDHJ0057 and Grant MZGC20240050. *(Corresponding author: Zhiliang Liu).*

Siyu You and Huayi Gou are with the School of Mechanical and Electrical Engineering, and also with the Glasgow College, University of Electronic Science and Technology of China, Chengdu, 611731, China (email: 2022190502023@std.uestc.edu.cn)

Leilei Yang and Zhiliang Liu are with the School of Mechanical and Electrical Engineering, University of Electronic Science and Technology of China, Chengdu, 611731, China (e-mail: zhiliang_liu@uestc.edu.cn)

Mingjian Zuo is with the School of Mechanical and Electrical Engineering, University of Electronic Science and Technology of China, Chengdu, 611731, China, the School of Qilu Transportation, Shandong University, Jinan, China, 250061, and also with the Qingdao Mingserve Tech. Ltd., Qingdao, 266041, China. (e-mail: zuo.mingjian@mingserve.com, ORCID: 0000-0002-8607-2923) the signal envelope, and the area under the envelope, to detect and quantify damage in SWRs [10]. Ren et al. present a novel shaking noise elimination method for detecting LFs in SWRs using MFL signals, which leverages a circular median filter to effectively distinguish and remove shaking noise while preserving the integrity of flaw signals [11]. Huang et al. introduce an adaptive fast Walsh-Hadamard transform method for extracting MFL signals of LFs under noisy conditions, utilizing adaptive particle swarm optimization to optimize transformation coefficients and enhance denoising performance [12]. Signal-based LF detection methods are relatively simple to implement and do not require complex sensor arrays, as they can use data from a limited number of sensors. This makes them cost-effective and easier to deploy in practical applications. Additionally, signal-based methods have well-established signal processing techniques and theoretical foundations, making them robust and reliable in many scenarios. However, they often struggle to fully capture multi-dimensional data, particularly in the case of complex or subtle LFs that are not easily represented by basic signal features. Furthermore, the accuracy of these methods may be limited by the quality of the signal, as noise can obscure subtle LF signatures.

Image-based methods diagnose LFs by detecting images generated from MFL raw signals by using image processing technology. For example, Wang et al. present a convolutional neural network (CNN)-Transformer-based model enhanced with transfer learning for LF detection, employing empirical mode decomposition for noise elimination and matrix reconstruction to transform MFL signals into images for analysis [13]. Yi et al. present an uncertainty-aware deep learning model for SWR defect detection, which employs the Gramian angular field to transform 1-D MFL signals into 2-D images, capturing spatial and temporal structures for enhanced defect identification [14]. Pan et al. utilize the object detection model YOLOv5 to detect defect areas in MFL images generated by interpolation [15]. However, Challenges such as the need for high-precision signal acquisition, high computational costs, large training datasets, and model interpretability remain significant hurdles for real-world applications of deep learning methods [16].

Traditional methods for analyzing MFL signals typically focus on mitigating noise to improve defect detection accuracy. However, the target-feature-oriented (TFO) method proposed by Pan et al. takes a novel approach by analyzing the characteristics inherent to the defects themselves [17]. Instead of simply removing noise from the entire signal, the TFO method focuses on the spatial features of the defect for denoising, using a defect-like template to extract and refine defect features in the MFL image. This approach emphasizes the extraction and optimization of defect-related features, allowing the preservation of spatial information and making it more effective for detecting defects even in noisy environments.

The TFO method achieves good performance in the localization of defects and image denoising. However, it still has some limitations:
1) The TFO method utilizes a single-scale Prewitt operator to match defect areas, which may not be effective for capturing defects of varying sizes and shapes. This could lead to missed detections or false positives, particularly for smaller or irregularly shaped defects.
2) The method does not dynamically adapt to changes in inspection speed or sampling frequency. This inflexibility can result in suboptimal performance when inspecting SWRs under varying conditions, as the defect signatures might change with different inspection speeds.

To overcome the limitations of existing methods, we propose an innovative adaptive spatial sampling resolution (SSR) TFO (AS-TFO) framework that dynamically adapts to changes in inspection speed and sensor sampling frequency. Our method utilizes a multi-scale pyramid structure to capture and enhance defect features based on working conditions. The main contributions of this article can be summarized as follows:

1) We analyze the characteristics of MFL images collected under different SSRs and discuss the limitations of existing single-scale TFO methods.
2) We design a multi-scale image pyramid structure that extracts features at various resolutions to overcome the limitations of single-scale methods. The proposed method ensures that both coarse and fine defect patterns are captured, enhancing the accuracy and robustness of defect detection.
3) We introduce adaptive convolution kernels and weighted image fusion mechanisms, which dynamically adjust the kernel size and pyramid layer weights according to the SSR scenario. This adaptive strategy ensures optimal feature fusion across scales, providing better performance under varying SSR scenarios.

This article is organized as follows: Section II outlines the generation process of MFL images and the Effect of SSR on MFL image quality, discussing the challenges posed by varying inspection speeds and sampling frequencies. In Section III, we detail our proposed AS-TFO framework, highlighting the construction of the image pyramid, the adaptive adjustment mechanisms, and the LF feature extraction process. Section IV presents the results of our method applied to experimental MFL data, demonstrating the effectiveness of our approach under varied SSRs. Section V provides an evaluation of the proposed adaptive mechanisms and comparative analysis with state-of-the-art (SOTA) methods. Finally, Section VI concludes the article, summarizing our contribution.

## II. Effect of SSR on MFL Image Characteristics and Denoising

In this section, the method of MFL image generation is introduced. Based on generated MFL images, the definition of SSR is illustrated and its effects on MFL images are comprehensively analyzed.

### A. MFL Image Generation

In the proposed method, we use multi-channel Hall sensors to collect MFL signals in the axial direction. The received axial signal is further transformed into an MFL image.

The generation process of the MFL image is depicted in Fig. 1. Initially, we perform detrending on the received *n*-channel signal (Fig. 1(a)) to standardize it to a common baseline, which is shown in Fig. 1(b). This step is crucial for emphasizing the low-frequency signal while mitigating the influence of any signal trends. The received MFL data is denoted as $x[m,n]$ for the $m$th sample value received in the $n$th channel, $m = 1, 2, …, M$ and $n = 1, 2, …, N$, where $M$ is sample number and $N$ is channel number. The detrending process can be defined as:

$$b[m, n] = \frac{1}{2L_a} \sum_{k=m-L_a}^{m+L_a-1} x[k, n] \quad (1)$$

where $x[k,n]$ is the trend signal, $2L_a$ is the span of the moving average filter, and $k$ serves as the filter's internal step index. The detrended signal $y[m,n]$ is then computed using the formula:

$$y[m, n] = x[m, n] - b[m, n] \quad (2)$$

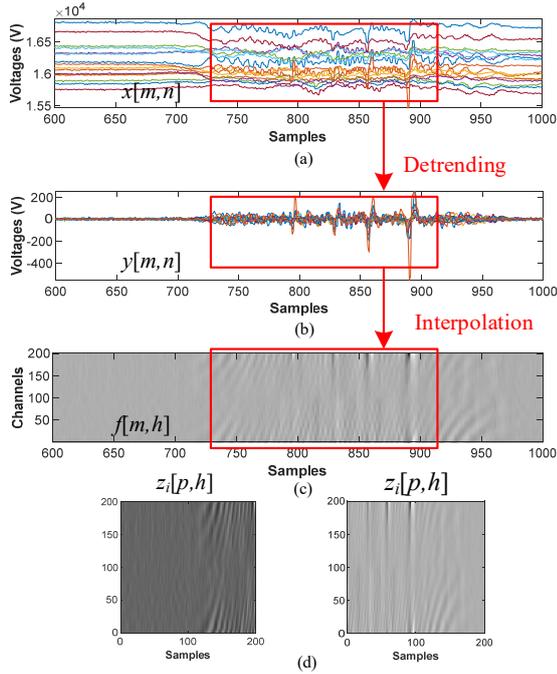

Fig. 1 MFL image generation process. (a) Raw MFL axial signals. (b) Detrended MFL signals. (c) MFL image after interpolation. (d) MFL images after segmentation.

After completing the detrending procedure, the signals from all 16 channels are combined to create a 16-width image. This image is then normalized to a range of -1 to 1. This normalization is essential because maintaining actual units during subsequent data processing could lead to alterations in the data's units, stripping the data of any meaningful physical significance. Following this normalization, the detrended signal is further interpolated along its channel dimension to enhance resolution. In this study, we utilize the spline interpolation method, known for its capability to produce smooth results and reduce artifacts and distortions in the data [18]. The resulting interpolated data, shown in Fig. 1(c), is denoted as $f[m,h]$, where $h=1,2,…,H$, and $H$ represents the height of the MFL image, which is set to 200 for this research. Following this, the interpolated data is divided into smaller segments shown in Fig. 1(d), which is defined as follows:

$$z_i[p, h] = f[p + (i-1) \times p, h] \quad (3)$$

where $z_i[p,h]$ is the $p$th sample of the $h$th channel inside the $i$th segment, $i=1, 2,…, [M/P]$, $p=1, 2, …, P$, $P$ is the length of each MFL image segment, and $P=200$ in this paper.

### B. Effect of Inspection Speed on MFL Image Quality

Inspection speed plays a crucial role in the quality of MFL images. One major impact is that as inspection speed increases, the number of sampling points collected over the same distance decreases. This reduction leads to an axial compression of data in the resulting image, which can cause defect features to appear blurred and indistinct. This effect is particularly problematic for identifying smaller or more complex LFs. Worth mentioning, another often overlooked aspect is the introduction of eddy current. As the sensor moves along the axial direction, it cuts through magnetic field lines, which can induce motion-induced eddy currents (MIECs). These MIECs typically form near the magnetic poles and LFs. At the magnetic poles, MIECs can interfere with the SWR's magnetization, resulting in an uneven distribution of the magnetic field. Near LF sites, MIECs disturb the symmetry of the magnetic leakage signal, causing distortions commonly referred to as the velocity effect [19], [20]. Such the velocity effect can distort strand noise and LF image area.

Fig. 2 illustrates MFL images captured at different inspection speeds. In the low-speed scenario (Fig. 2(a)), the MFL images exhibit clear and well-defined LF features. As the inspection speed increases, as seen in Fig. 2(b) and Fig. 2(c), the LFs begin to exhibit noticeable compression in the axial direction. Simultaneously, the slope of the strand noise also increases with the rising speed. In Fig. 2(c), the highlighted area indicates strand noise regions affected by motion-induced distortion. The affected region of the strand wave has similar characteristics to the defective portion, with a bright rectangular stripe sandwiched by dark stripes in the adjacent regions on either side along the axial direction.

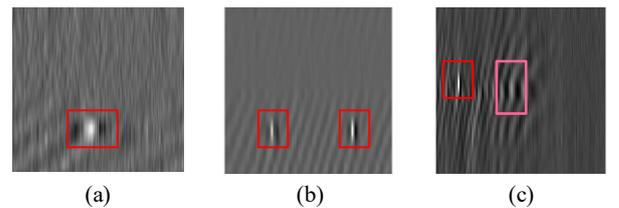

Fig. 2 The shapes of the same LF under different inspection speeds. (a) Low speed scenario (0.1m/s). (b) Medium speed scenario (0.5 m/s). (c) High speed scenario (1.2 m/s).

### C. Effect of Sensor Sampling Rate on MFL Image

The sensor sampling rate, or the frequency at which the sensor collects data points, also significantly affects MFL image characteristics. A high sampling rate allows the sensor to capture more data points per unit length of the rope, producing images with greater resolution and detail. Conversely, a low sampling rate may result in sparse data, making it difficult to identify finer LF characteristics. Different sampling rates yield images with distinct characteristics, as illustrated in Fig. 3.

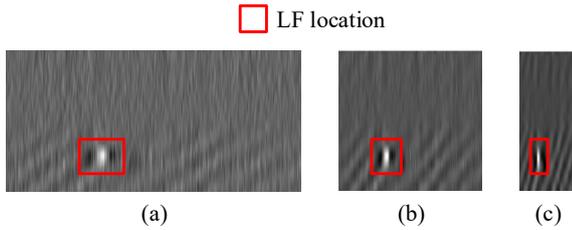

Fig. 3  MFL images under the three sensor sampling rates. (a) 250 Hz (b) Downsampled to 125 Hz. (c) Downsampled to 50 Hz.

Fig. 3(a) displays an MFL image under the original sensor sampling rate, which is 250 Hz. Then, the raw data is downsampled to 125 Hz and 50 Hz and is generated into MFL images shown in Fig. 3(b) and Fig. 3(c). From these figures, it can be noted that as the sampling rate decreases, the number of data points captured per unit time reduces, resulting in axial compression of LF features within the image. This compression effect causes LFs to appear narrower along the axial direction, and it also increases the slope of strand noise, complicating the distinction between LFs and strand noise. Notably, this phenomenon mirrors the effect of increased inspection speed, where higher speeds similarly compress LF representation in the axial direction and steepen strand noise patterns.

### D. Effect of SSR and Its Role

From the analysis above, both the sensor sampling rate and inspection speed have homogeneous effects on MFL images. This is due to that they both affect the information that the sensor can obtain in unit distance. In the realm of remote sensing, spatial resolution is a critical concept that defines the capability of the imaging system to differentiate between closely spaced objects or features [21], [22]. High spatial resolution allows for the detection of finer details, whereas lower spatial resolution may result in the loss of subtle features, thus affecting the accuracy of analysis [23]. Building on the definition of spatial resolution in remote sensing, in our context, we define SSR as the density of information captured per unit length in an axial MFL signal. SSR can be calculated as:

$$f_{spatial} = \frac{f_s}{v} \quad (4)$$

where $f_{spatial}$ represents the SSR, $f_s$ and $v$ refer to the sensor sampling rate and inspection speed respectively. This metric reflects the density of information captured in MFL images. At high SSR scenarios, LFs are distinctly represented with prominent axial dimensions, aiding in accurate identification, as shown in Fig. 4(a).

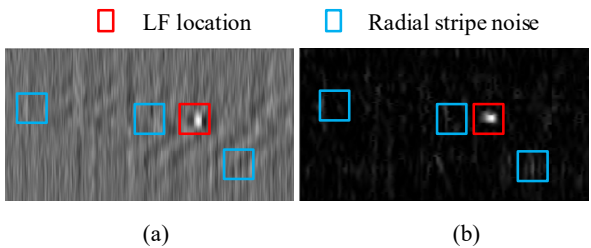

Fig. 4  MFL image under a high SSR scenario. (a) Raw image. (b) Image after using the single-scale TFO denoising method.

However, the presence of significant radial stripe noise can complicate the analysis. This noise can mimic LF characteristics, resulting in small-scale convolution templates misidentifying these artifacts as LFs as shown in Fig. 4(b).

In low SSR scenarios, image quality deteriorates due to compression of LFs in the axial direction and loss of critical details. The pixel density becomes insufficient to capture the full extent of LFs, causing them to appear as thin lines. Additionally, as shown in Fig. 5, the strand wave areas with steep slopes have similar features to LF areas. Such features can obscure the detection of actual LFs if larger convolution templates are used, increasing the risk of false positives in a single-scale TFO framework.

TABLE I
SUMMARY OF MFL IMAGE FEATURES IN DIFFERENT SSR SCENARIOS

| SSR Scenario | Stripe Noise | Strand Noise | LF Area |
|---|---|---|---|
| Low | None | High slope, similar to LFs | Compressed in the axial direction, difficult to identify small LFs |
| High | Small scale, similar to LFs | Proper slope | Stretched in the axial direction, more defined but with distorted stripes |
| Optimal | None | Proper slope | Clear, well-defined LF areas with minimal distortion |

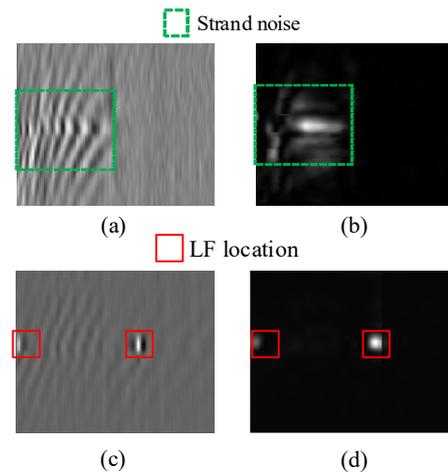

Fig. 5  Raw MFL image and single-scale TFO denoised image. (a) Raw image under a high SSR scenario. (b) Image after using single-scale TFO denoising under a high SSR scenario. (c) Raw image under a low SSR scenario. (d) Image after using single-scale TFO denoising under a low SSR scenario.

An optimal SSR strikes a balance between clarity and noise, significantly reducing misleading artifacts while maintaining clear LF representation. As shown in Fig. 5(c), the LF image is clear and the axial size of LF is proper. Compared with the MFL image in low SSR scenarios, there is almost no radial stripe noise in the image and the LF is clearer. Fig. 5(d) shows the result after single-scale TFO denoising. It can be found that no strand noise remains after single-scale TFO denoising and all LF areas are retained.

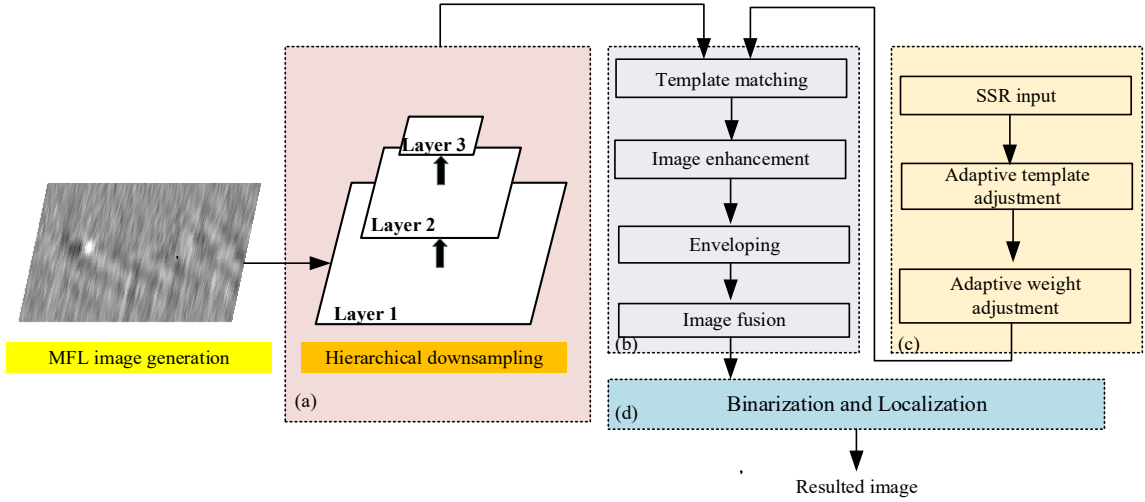

Fig. 6 The overview of the proposed method. (a) The construction of an image pyramid. (b) The process of image feature extraction. (c) The proposed SSR adaptive adjustment mechanism. (d) The binarization and localization of the MFL image.

An optimal SSR is the ideal condition for MFL image-based methods. However, consistently achieving stable optimal SSR in real-world operational conditions is challenging. Consequently, single-scale TFO denoising may be inadequate for effectively processing MFL images across varying SSRs.

## III. METHODOLOGY

### A. Multi-Resolution Image Pyramid Construction

As mentioned in the previous section, in a low SSR scenario, the size of the LF area is small and hard to detect. Meanwhile, despite the clarity of LF image in high SSR scenario, some radial stripe noise with similar characteristics to LF may cause false negatives. To handle various LF scales, we construct a Gaussian pyramid of MFL images with three layers, each representing a different resolution. Such an image pyramid mechanism is widely used in CNNs to enhance the ability to process images in different scales [24], [25], [26]. Starting from the original MFL image, we iteratively downsample each subsequent layer by a factor of 0.5 to create a specified number of levels. As for the $j$th layer of the inputted image, denoted as $z_{ij}[p_j,h_j]$, this downsampling process can be expressed as follows:

$$Z_{ij+1}[p_{j+1},h_{j+1}] = \frac{1}{\sigma^2}\sum_{m=0}^{\sigma-1}\sum_{n=0}^{\sigma-1} Z_{ij}[\sigma p_j + m, \sigma h_j + n] \quad (5)$$

where $j$=1, 2 and $\sigma$ is the downsampling scaler, which is 0.5 in this article. This downsampling process is hierarchical and it is demonstrated in Fig. 6(b).

### B. MFL Image Denoising and LF Localization

#### (1) Adaptive Template Matching

The LF signal has an important feature in that it has a combination of a peak and valley in the axial direction [15]. This feature demonstrated in the MFL image, can be summarized as an adjacent combination of a dark area and a bright area. Therefore, a template is designed to extract the LF feature in the MFL image, shown in Fig. 7(a). The areas filled with −1 and 1 represent the dark and bright areas in the LF area respectively. By doing a convolution operation on the MFL image, we can obtain the template response values of areas in MFL images, representing the similarity between the LF template and possible LF regions.

In the proposed method, we apply an LF template on each layer of the image pyramid to extract LF features in different image scales. As for an MFL image segment $z_{ij}[p_j,h_j]$, the convolution operation involves sliding the convolution kernel $K_a$ over the image, computing the weighted sum of pixel intensities within the kernel region at each position:

$$C_j(p,h) = \sum_{a=-\frac{K_a}{2}}^{\frac{K_a}{2}} \sum_{b=-\frac{K_a}{2}}^{\frac{K_a}{2}} K_a(a,b) \cdot z_{ij}(p-a,h-b) \quad (6)$$

Here, $C_j(p,h)$ represents the output value after applying the convolution at pixel $(p,h)$ in layer $j$, reflecting the extracted feature at that location. The term $K_a$ denotes the weights of the convolution kernel at position $(a,b)$, defining the importance of each pixel in the region covered by the kernel, while $z_{ij}[p-a,h-b]$ corresponds to the pixel value at position $(p−a,h−b)$ in the image.

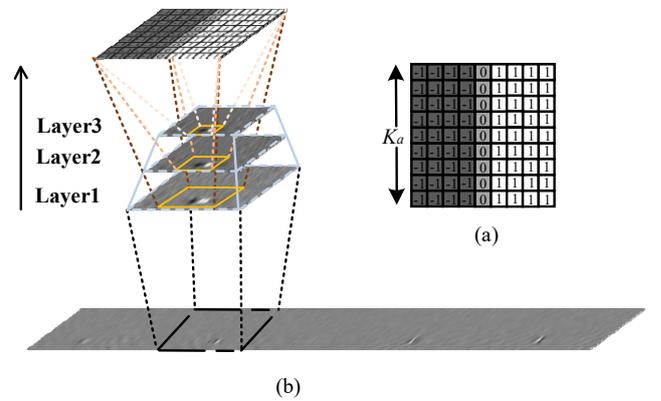

Fig. 7 The process of template matching. (a) The shape of the convolution template. (b) Process of pyramid downsampling and template matching.

As for the LF image area in low SSR scenarios, its size is relatively small so it should be matched by a relatively small template. Conversely, images in high SSR scenarios should be matched by a large template. Therefore, we introduce an SSR compensation mechanism for convolution kernel size adjustment. Firstly, the SSR is normalized. The SSR is first calculated by (4). Then, one extremely low SSR is chosen as the extreme reference, where the extreme sensor sampling rate $f_{s\_extreme}$ is 250 Hz and extreme velocity $v_{extreme}$ is 1.5 m/s, and the extreme SSR $f_{spatial\_extreme}$ can be calculated as a normalization parameter and it is the threshold of adaptive mechanism. This choice $f_{s\_extreme}$ and $v_{extreme}$ is based on the practical limitations of our experimental setup and the need to define a clear reference point for normalization. It is important to note that if the sensor's sampling rate changes, the corresponding extreme speed would also change, but the extreme SSR value would remain consistent, ensuring the generalizability of the proposed method across different operational conditions. The detailed experimental setup, including the rationale for selecting these specific sensor parameters, will be further elaborated in Section IV. Once the SSR exceeds this threshold, we begin implementing our adaptive convolution kernel adjustment mechanism. By setting the extreme condition, the SSR can be normalized by the equation:

$$\mu_{spatial} = \frac{v}{v_{extreme}} \cdot \frac{f_{s\_extreme}}{f_s} = \frac{f_{spatial\_extreme}}{f_{spatial}} \quad (7)$$

where $\mu_{spatial}$ is the normalization value of SSR. After that, a basic kernel size is chosen and it is as small as possible to detect fine LF area in MFL image. Therefore, this article uses both theoretical calculation and measuring MFL images for choosing a basic convolution size. Supposing that for an LF with 20 mm axial length, under high-speed conditions, with a sampling rate of 250 Hz and a speed of 1.5 m/s, the magnetic field around the LF can be expected to represent the LF's length. According to the definition of SSR given by (4), the sampling points $N_s$ within a given axial distance $d$ can be calculated as follows:

$$N_s = f_{spatial} \times d \quad (8)$$

Therefore, an LF of 20 mm axial length would correspond to approximately 5 sampling points, meaning this LF would appear as an axial length of 5 pixels on the image.

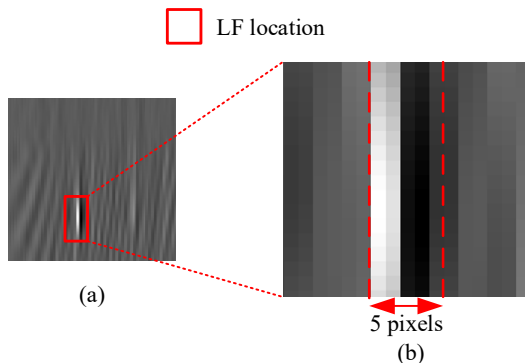

Fig. 8. MFL image under extremely low SSR scenario. (a) Raw MFL image (b) Magnified LF area.

For further validation, we conducted actual image validation, as shown in Fig. 8(b), where under extremely low SSR scenarios, the axial dimension of the LF image is indeed approximately 5 pixels, which aligns with our calculations. For scenarios where the SSR is relatively low, the basic convolution kernel provides sufficient sensitivity to detect LFs accurately.

Then, a kernel size compensation mechanism is proposed. This mechanism is governed by a sensitivity factor $\alpha$, which determines the extent to which the convolution kernel size is adjusted in response to changes in the SSR. The sensitivity factor essentially controls how effectively the convolution kernel adapts to variations in sampling rate. The adaptive mechanism of the convolution kernel can be expressed by:

$$K_a = \lceil K_{base} + \alpha \cdot (1 - \mu_{spatial}) \rceil \quad (9)$$

where the $K_a$ is the kernel size after adaptive adjustment and $K_{base}$ is the size of the base convolution. From (9), the adaptive kernel size increases as $\mu_{spatial}$ decreases, which aligns with the demand for larger template size as SSR increases. The choice of sensitivity factor is based on extensive empirical validation. Through evaluating various SSR scenarios with different sensitivity factors, we found that setting the sensitivity factor to 5 achieves the best performance under various SSR scenarios.

After convolutional template matching, to manage boundary effects caused by convolution near the edges of the image, padding is applied to the boundary regions. In particular, the pixel values in the leftmost and rightmost columns are adjusted by replicating the values of the nearest interior columns, which helps maintain consistency in the convolution output across the entire image [27]. After convolution, the absolute values are taken to convert negative intensities to positive values, which simplifies subsequent processing steps.

*(2) Gamma Transformation*

Following template matching, a gamma transformation is applied to enhance contrast in the detected features. Let γ represent the gamma transformation parameter, the enhanced image $E_{ij}(p,h)$ at layer $j$ is given by:

$$E_{ij}(p,h) = |C_{ij}(p,h)|^{\gamma} \quad (10)$$

This nonlinear transformation accentuates differences in pixel intensities. It allows for improved visualization and feature enhancement, particularly in low-intensity regions.

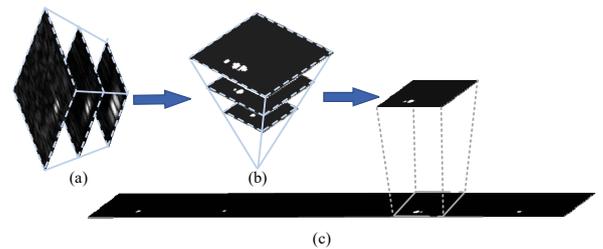

Fig. 9 The process of image denoising. (a) Results of enveloping in each layer. (b) Process of image feature fusion. (c) Results of binarization.

*(3) Enveloping*

After gamma enhancement, an envelope detection method is applied to the result. This process involves interpolating and identifying peaks in the gamma-transformed image $E_{ij}(p,h)$, capturing the envelope of intensity variations across each pixel column. This envelope, denoted as $F_{ij}(p,h)$, represents the final feature-enhanced image at each layer, highlighting MFL signal variations that indicate potential LFs.

*(4) Adaptive Weighted Image Fusion Mechanism*

In the previous steps, images in each layer in the pyramid are processed. As for those images under high SSR scenarios, they should be inspected in a high-resolution layer. Conversely, images under low SSR scenarios should be inspected in a low-resolution layer. To further refine feature extraction, this article designs a dynamic fusion mechanism to adjust the weights of different layers based on the SSR. Firstly, the weights of different layers are calculated based on the normalized SSR. This weight adjustment calculation process can be expressed as follows:

$$w_1 = \mu_{spatial}^2 \tag{11}$$

$$w_2 = 2 \cdot \mu_{spatial} \cdot (1 - \mu_{spatial}) \tag{12}$$

$$w_3 = (1 - \mu_{spatial})^2 \tag{13}$$

where $w_1$, $w_2$, $w_3$ are weights of the high-resolution layer, the medium-resolution layer, and the low-resolution layer respectively. This weighting mechanism ensures that features most relevant to the current operational conditions are emphasized.

After calculating the weights of each layer, images in all layers are upsampled to the original resolution and fused into a single image based on weights for further processing. This feature fusion mechanism can be expressed by:

$$F_{ij}^{final} = w_j \cdot F_{ij} + (1 - w_j) \cdot F_{ij+1}^{final} \tag{14}$$

where the weight $w_j$ determines the contribution of the current layer $F_{ij}$, while $(1 - w_j)$ represents the contribution from the previously fused layers $F_{ij}^{final}$.

*(5) Thresholding and Binarization for LF Segmentation*

In this study, we adopt the adaptive thresholding and binarization method from [17] to segment LFs in the enhanced MFL images. The adaptive thresholding process analyzes the number of disjoint regions in the binarized image across a range of threshold values. This helps identify the optimal threshold that best separates LF signals from background noise. This thresholding approach ensures robust LF localization even in the presence of noise. Afterward, the binarization step converts the enhanced MFL image into a binary format, where LF regions are marked white and non-defective areas are black, effectively isolating the LFs for further analysis.

## IV. CASE STUDY

To validate the proposed method, a case study is conducted. The experimental procedure is first introduced and results of AS-TFO denoising and LF localization are then demonstrated.

### A. Experimental Setup

Based on the MFL principle, the MFL sensor shown in Fig. 10 is designed and utilized to collect MFL signals. The permanent magnets NdFeB are used for magnetizing the SWR and formulating a magnetic circuit between the armatures made of soft irons and the SWR. The Hall sensors (model type: A1301, sensitivity: 2.5mV/Gauss), which can transform magnetic signals into electrical signals, are used to detect MFL signals in the axial direction. To obtain circumstantial information on the SWR, 16 Hall sensors are uniformly arranged in a ring, which composes a uniform circular array. The inner diameter of the sensor is 40 mm. The sensor can be applied under two modes, which are portable mode and online mode respectively [28]. This article uses portable mode in all experiments.

To obtain data from both high SSR and low SSR scenarios, experiments under different inspection speeds are performed. The sensor's sampling rate is 250 Hz. In this article, two ranges of high and low SSR are: low (sampling rate: 250 Hz, inspection speed: 0~0.5 m/s), and high (sampling rate: 250 Hz, inspection speed: 1~1.5 m/s). The experiments are performed on a SWR with 4 external LFs, and the sizes of these 3 LFs are 1 broken wire, 2 broken wires, and 3 broken wires respectively. Such variation provides a range of scenarios to test the robustness of the proposed method. The diameter of the SWR is 32mm. Meanwhile, the dataset is collected across four different sensor positions. These four positions represent distinct orientations of the sensor relative to the defect, spaced at 90-degree intervals, ensuring comprehensive coverage of the defect's spatial characteristics from multiple angles. In each set of data, the detector is repeatedly moved along the axial direction of the SWR two times. Additionally, we use a camera to record videos of each experiment at 60 Hz. By counting the number of video frames that pass over the LF areas and knowing the length of the SWR, we can calculate the inspection speed.

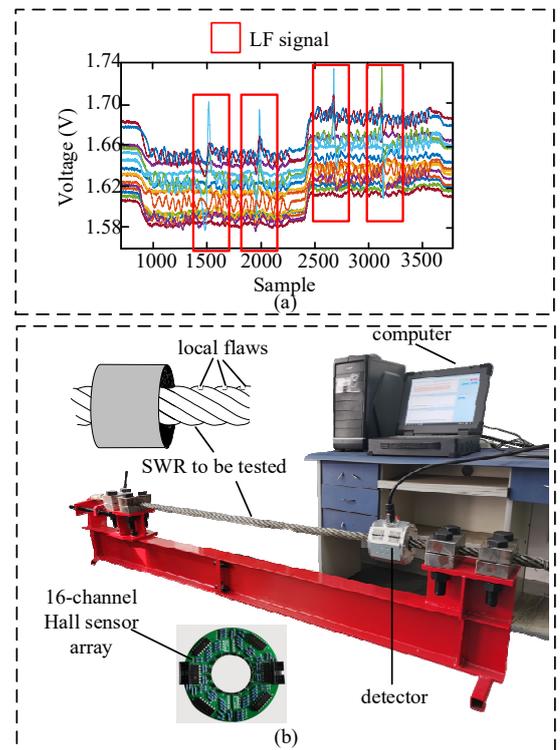

Fig. 10 MFL data acquisition system (a) An axial MFL data collected by 16-channel Hall sensor (b) Components of MFL data acquisition system

### B. Results of Image Denoising and LF Localization

The method used in this step constructs the top two layers of the pyramid. The original image undergoes two iterations of down-sampling, resulting in two reduced-resolution images. Fig. 11(a) and Fig. 11(b) display MFL images in both low SSR and high SSR scenarios respectively. Before images are denoised in each layer in a pyramid, the adaptive convolution size and layer weight are calculated based on the SSR of the corresponding image. Then, images in each layer are denoised through template matching.

From Fig. 11(e), in the low SSR image layer, stripe noise remains in the high-resolution layer due to the similarity between small-scale LFs and stripe noise. However, only LF areas are effectively matched in the low-resolution layer of the image pyramid, validating the effectiveness of the pyramid mechanism in reducing stripe noise. In contrast, Fig. 11(f) shows that in a high SSR scenario, the LF area is matched properly in a high-resolution layer with little noise being recognized. However, in the low-resolution layer, many strand noise areas are recognized due to their high slope and similarity with LF areas.

After template matching, features in the image pyramid are enhanced and enveloped, then fused by an adaptive layer weight mechanism. As shown in Fig. 11(g) and Fig. 11(h), after feature fusion, images under different SSRs are denoised, leaving only LF areas. This can be attributed to the adaptive layer weight mechanism. The adaptive layer weight mechanism reduces the weight of the high-resolution layer in high SSR scenarios. This helps eliminate the effects of remaining stripe noise. In the low SSR scenario, the weight of the low-resolution layer is reduced, leaving only LF areas matched in the high-resolution layer remaining in the final fused image. After that, the LF areas are then localized in fused images, as shown in Fig. 11(i) and Fig. 11(j). The final output image demonstrates that all LFs are recognized correctly and there are no non-defect areas that are mismatched as LF areas. This confirms the accuracy and effectiveness of the denoising and LF localization process.

As a result, the proposed AS-TFO mechanism not only improves the performance of image denoising by image pyramid and adaptive mechanisms but also enhances the overall robustness of LF localization by adaptive threshold and binarization, making it more robust to noise and more accurate in identifying LFs, regardless of SSR scenarios.

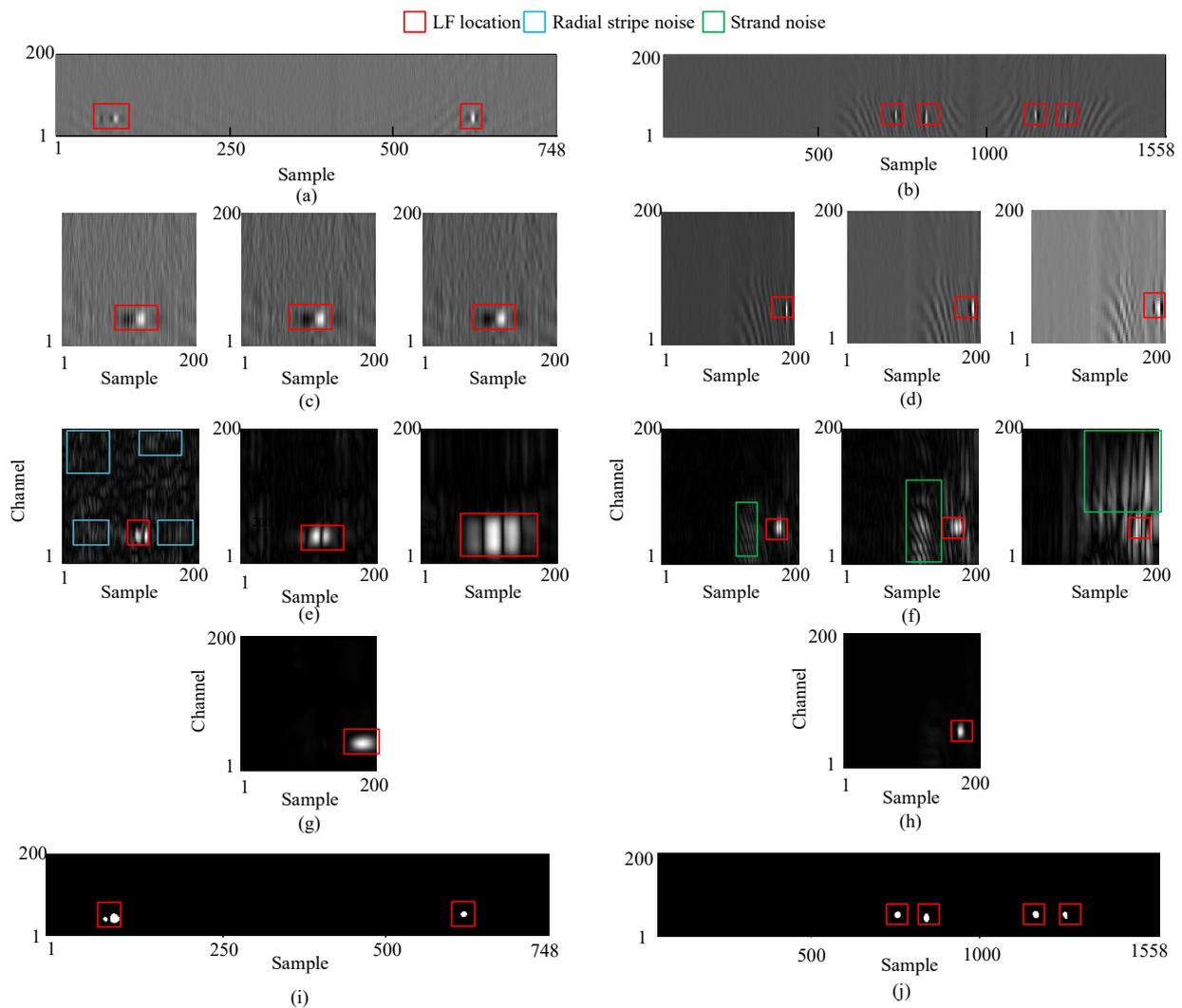

Fig. 11 Experimental MFL image. (a) Raw image under a high SSR scenario (b) Raw image under a low SSR scenario (c)-(d) MFL images after downsampling. (e)-(f) MFL images after template matching (g)-(h) MFL image after feature fusion. (i)-(j) MFL images after thresholding and binarization.

## V. Performance Evaluation and Comparison

To evaluate the effectiveness of the proposed AS-TFO method, a comprehensive performance comparison was conducted in this section.

### A. Effectiveness Evaluation of Adaptive Mechanism

To evaluate the effectiveness of the adaptive mechanism, performance comparisons are conducted against both single-scale TFO and pyramid multi-scale without adaptive weighted fusion methods. The comparison is conducted on 50 sets of high SSR data and 50 sets of low SSR data. The results are summarized in TABLE II and TABLE III, with precision, recall, and F1 score as key metrics. Precision indicates the proportion of true positives (TP) among all positive detections, assessing detection accuracy. Recall measures the proportion of actual positives correctly identified, indicating detection completeness. The F1 score, as the harmonic mean of precision and recall, provides an overall performance balance.

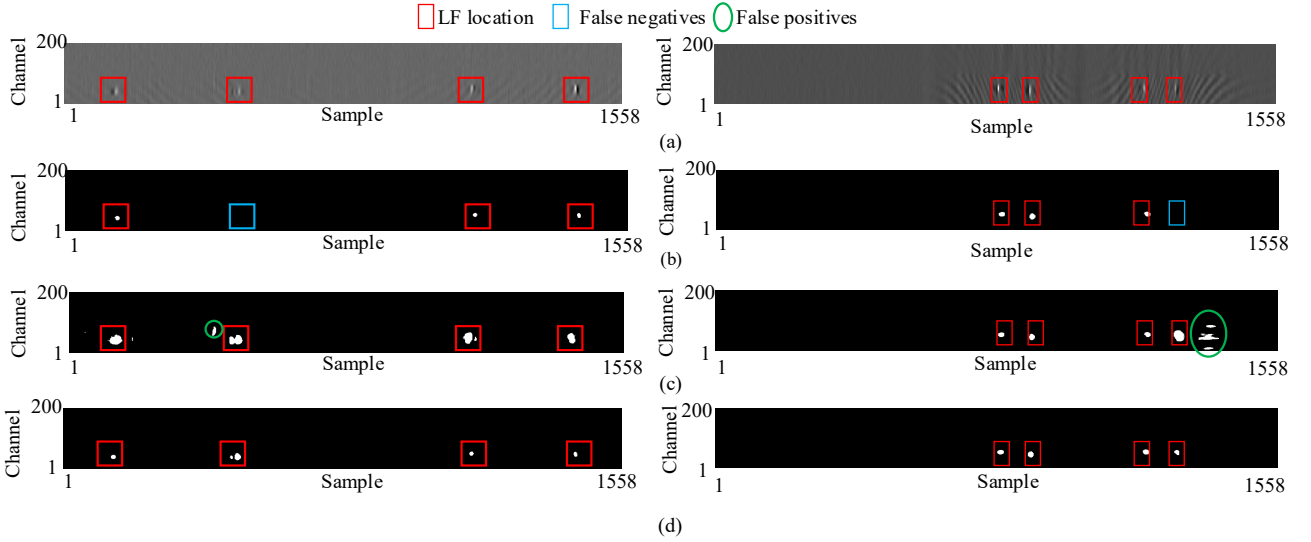

Fig. 12 Localization performance comparison under a high SSR scenario (left) and a low SSR scenario (right). (a) Raw MFL images. (b) Localization results of single scale TFO method. (c) Localization results of multi-scale pyramid TFO without adaptive weighting. (d) Localization results of the proposed AS-TFO method.

TABLE II
PERFORMANCE COMPARISON UNDER THE HIGH SSR SCENARIO

| Performance Metrics | Single-scale TFO [17] | Unweighted multi-scale TFO | The proposed method |
|---|---|---|---|
| TP | 53 | 60 | 62 |
| FP | 5 | 15 | 5 |
| FN | 10 | 3 | 1 |
| Precision | 91.38% | 80.00% | **92.54%** |
| Recall | 84.13% | 95.24% | **98.41%** |
| F1 score | 87.61% | 86.96% | **95.38%** |

TABLE III
PERFORMANCE COMPARISON UNDER THE LOW SSR SCENARIO

| Performance Metrics | Single-scale TFO [17] | Unweighted multi-scale TFO | The proposed method |
|---|---|---|---|
| TP | 127 | 141 | 148 |
| FP | 21 | 34 | 8 |
| FN | 25 | 11 | 4 |
| Precision | 85.81% | 80.57% | **94.87%** |
| Recall | 83.55% | 92.76% | **97.37%** |
| F1 score | 84.67% | 86.24% | **96.10%** |

In both tables, the proposed method demonstrates superior localization accuracy. The F1 score of the proposed method is 95.38% and 96.10%, which are higher than those of the single-scale TFO method (87.61% and 84.67%) and unweighted Multi-scale TFO method (86.96% and 86.24%), indicating the greater performance of the proposed methods under both high and low SSR scenarios. Some results, illustrated in Fig. 12, demonstrate the final processed images and the original unprocessed ones. From Fig. 12(c) and Fig. 12(d), both multi-scale methods demonstrate superior LF localization effectiveness, as the single-scale TFO method has lower accuracies in both low SSR and high SSR scenarios. This is due to that single-scale TFO only extracts features on a limited scale and does not consider the variations of LF image patterns in various SSRs. The proposed pyramid mechanism and adaptive convolution mechanism address this issue, extracting LF features in different scales. Additionally, though the non-weighted multiscale approach shows a higher recall compared to the single-scale method, it shows a much higher false positive (FP) rate, making its precision even worse. This can be ascribed to that multi-scale method without weighting fuse all extracted features from all layers without any filtering, making some stripe noise and strand noise mismatched, as shown in Fig. 12(c). However, with our adaptive weighted fusion mechanism, such redundant features can be eliminated and LF features can still be retained, significantly improving outcomes with fewer false negatives (FN) and higher precision.

While the proposed adaptive mechanism demonstrates significant improvements in defect detection accuracy and robustness, it can not eliminate certain sorts of noise and may still cause few FP. For example, as shown in Fig. 13(b), one non-defect region is still misclassified as LF.

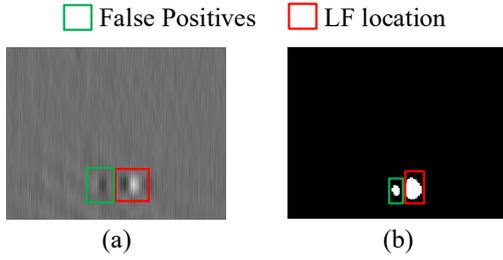

Fig. 13 One example of a false positive. (a) Original MFL image. (b) Resulted image after AS-TFO denoising

Noise such as strand noise and stripe noise can be effectively eliminated using template matching methods. However, as observed in Fig. 13(a), a mismatched region corresponding to a dark rectangular noise area in the original image exhibits morphological characteristics similar to defects. This noise likely arises from signal distortion caused by a rare combination of jitter, strand waves, and electrical noise during the experiment, which alters its inherent signal features, making it resemble LF but with a smaller amplitude. Consequently, this residual area shares traits with LF and is not fully filtered out during the adaptive template matching stage. Additionally, under high SSR conditions, the increased information density amplifies the scale of such noise in the images, which explains why precision under high SSR conditions (92.54%) is slightly lower than under low SSR conditions (94.87%). Although this special case results in a small number of false positives, its impact on our method is limited, and further optimization of sensor stability and electrical noise could mitigate these rare distortions.

### B. Evaluation of Method Adaptability

To further evaluate the overall performance of the proposed method under various sampling conditions, we also conducted experiments on 50 sets of optimal SSR data. The overall results are summarized in TABLE IV.

TABLE IV
THE PERFORMANCE OF THE PROPOSED METHOD ON 32MM SWR DATASET

| Performance Metrics | Low SSR | High SSR | Optimal SSR | Overall |
|---|---|---|---|---|
| TP | 148 | 62 | 149 | 359 |
| FP | 8 | 5 | 2 | 15 |
| FN | 4 | 1 | 1 | 6 |
| Precision | 94.87% | 92.54% | 98.67% | 95.99% |
| Recall | 97.37% | 98.41% | 99.33% | 98.36% |
| F1 score | 96.10% | 95.38% | 99.00% | 97.16% |

Furthermore, to examine the generability and robustness of the proposed method under different LF types and SWR diameters, we also conducted experiments on another SWR with three internal LFs. The diameter of this SWR is 21mm and the diameters of these three internal LFs are 1 broken wire, 2 broken wires, and 3 broken wires respectively. All experiment procedures are consistent with the 32 mm SWR dataset. In each sort of SSR, 20 sets of data are collected. The results of the 21 mm SWR data set are shown in TABLE V.

As illustrated in TABLE V, the results show that the proposed method maintains strong performance, with a precision of 93.20%, recall of 96.00%, and an F1 score of 94.58%. Although the performance slightly decreases compared to the 32 mm SWR dataset, particularly in high SSR scenarios (Precision: 91.42% vs. 92.54%), the method still demonstrates high accuracy and robustness, especially in optimal SSR scenarios (Precision: 96.77%, Recall: 96.77%, F1 Score: 96.77%). The slight performance drop may be attributed to the increased difficulty of detecting internal LFs and the influence of smaller SWR diameters on MFL signal distribution. These results further validate the adaptability of the proposed method across different SWR structures and defect types.

TABLE V
THE PERFORMANCE OF THE PROPOSED METHOD ON 21MM SWR DATASET

| Performance Metrics | Low SSR | High SSR | Optimal SSR | Overall |
|---|---|---|---|---|
| TP | 34 | 32 | 30 | 96 |
| FP | 3 | 2 | 1 | 7 |
| FN | 1 | 2 | 1 | 4 |
| Precision | 91.89% | 91.42% | 96.77% | 93.20% |
| Recall | 97.14% | 94.11% | 96.77% | 96.00% |
| F1 score | 94.44% | 92.75% | 96.77% | 94.58% |

### C. Overall Performance Comparison

Based on the results of the two datasets, the overall performance of the proposed method is summarized in TABLE VI. To further validate the superiority of the AS-TFO, the performance of the proposed method is compared with the SOTA methods, summarized in TABLE VII.

TABLE VI
THE OVERALL PERFORMANCE OF THE PROPOSED METHOD

| Performance Metrics | Low SSR | High SSR | Optimal SSR | Overall |
|---|---|---|---|---|
| TP | 182 | 180 | 179 | 541 |
| FP | 11 | 10 | 3 | 24 |
| FN | 5 | 6 | 2 | 13 |
| Precision | 94.30% | 94.73% | 98.35% | 95.75% |
| Recall | 97.32% | 96.77% | 98.89% | 97.65% |
| F1 score | 95.79% | 95.73% | 98.61% | 96.69% |

TABLE VII
PERFORMANCE COMPARISON WITH SOTA METHODS

| Method | Name | Precision | Recall | F1 score |
|---|---|---|---|---|
| Pan et al. [17] | Single-scale TFO | 86.07% | 94.68% | 90.17% |
| Ren et al. [29] | Constant threshold | 29.89% | 92.86% | 45.22% |
| Pan et al. [15] | YOLOv5s | **96.31%** | 96.40% | 96.36% |
| Proposed Method | AS-TFO | 95.75% | **97.65%** | **96.69%** |

The performance comparison with SOTA methods reveals that the proposed AS-TFO outperforms the traditional constant threshold method, single-scale TFO method, and deep learning-based YOLOv5s, achieving SOTA performance. Specifically, our method achieves a precision of 95.75%, a recall of 97.65%, and an F1 score of 96.69%, significantly surpassing the results of the single-scale TFO method (86.07% precision, 94.68% recall, F1 score of 90.17%) and the constant threshold method (29.89% precision, 92.86% recall, F1 score of 45.22%). Meanwhile, the performance of the proposed method outperforms than deep learning-based method, achieving 0.33% higher F1 score than the AS-TFO method. These results highlight the effectiveness of the AS-TFO method, demonstrating its superior capability in handling various sampling conditions and achieving reliable detection performance without the need for training data.

8　　　　　　　　　　　　　　　　　　　　　　　　　　　　　　　　　　　　IEEE SENSORS JOURNAL, VOL. XX, NO. XX, MONTH X, XXXX## D. Evaluation of Computational Efficiency

While the proposed AS-TFO method demonstrates significant improvements in detection accuracy, it is essential to evaluate its computational efficiency and real-time feasibility for industrial applications. Therefore, we further conduct experiments to measure its processing time for a single image and frame per second (FPS). Moreover, to evaluate the demand for memory of the proposed method, the peak-value memory occupied by the program is recorded to represent the maximum demand for memory. Our experimental code is compiled and run on MATLAB 2023b. The experiments are performed on an office laptop equipped with a Ryzen 7 7840HS CPU and 32 GB of RAM. All metrics are tested under 5 different sets of data and averaged. The experimental results are presented in TABLE VIII.

TABLE VIII
EVALUATION OF COMPUTATIONAL COMPLEXITY

| Method | Processing time per image (ms) | FPS | Peak-value RAM (KB) |
|---|---|---|---|
| AS-TFO | 300.8 | 3.32 | 2240 |

As shown in TABLE VIII, with a sensor data acquisition rate of 250Hz per second, our method is capable of processing 3.32 images per second, each consisting of 200 data points. This results in a total processing capacity of approximately 664 data points per second, which exceeds the sensor's data acquisition rate, ensuring that the system can efficiently handle the incoming data stream without any bottlenecks. Furthermore, the peak memory usage of the method is 2240KB, indicating that the approach operates with relatively low memory requirements, making it suitable for deployment on embedded systems with limited resources.

## VI. CONCLUSION

This study investigated the SSR effect on MFL imaging for SWR inspection, with a focus on how sensor sampling rate and inspection speed influence image quality and LF detection accuracy. To mitigate these impacts brought by varying SSRs, we proposed an AS-TFO method that combines the image pyramid mechanism and adaptive layer weight and convolution mechanisms, enhancing detection robustness across different sampling conditions. The image pyramid mechanism and adaptive convolution size mechanism enhance the ability to extract LF features. The adaptive layer weight mechanism filters redundant features based on the corresponding SSR. Experimental results validate the effectiveness of the proposed method across different SSRs. Specifically, in our 32mm SWR dataset, under high SSR scenarios, our method achieved a precision of 92.54%, a recall of 98.41%, and an F1 score of 95.38%, outperforming traditional single-scale and unweighted multi-scale TFO methods. At low SSR scenarios, the proposed method maintained strong performance with a precision of 94.87%, recall of 97.37%, and an F1 score of 96.10%, demonstrating its adaptability and reliability. Overall, the proposed method has a precision of 95.75%, recall of 97.65%, and an F1 score of 96.69%, reaching SOTA performance among existing methods. Furthermore, computational efficiency tests demonstrate that the proposed method meets the requirements for real-time industrial applications.

Overall, the results of this study demonstrate that the proposed method effectively reduces issues related to LF compression and noise interference, ensuring higher robustness and accuracy across various operational conditions. These advancements provide a solid foundation for high-accuracy monitoring of SWRs in industrial applications, thereby enhancing the safety and reliability of critical infrastructure.

REFERENCES

[1] P. Zhou, G. Zhou, Z. Zhu, Z. He, X. Ding, and C. Tang, "A Review of Non-Destructive Damage Detection Methods for Steel Wire Ropes," *Applied Sciences*, vol. 9, no. 13, p. 2771, Jul. 2019, doi: https://doi.org/10.3390/app9132771.

[2] S. Liu, Y. Sun, X. Jiang, and Y. Kang, "A Review of Wire Rope Detection Methods, Sensors and Signal Processing Techniques," *Journal of Nondestructive Evaluation*, vol. 39, no. 4, Nov. 2020, doi: https://doi.org/10.1007/s10921-020-00732-y.

[3] Z. Hu, M. Dong, F. Jia, and E. Wang, "Fatigue damage evaluation and verification of steel wire rope based on magnetic flux leakage," *Journal of civil structural health monitoring*, vol. 14, no. 1, pp. 223–235, Oct. 2023, doi: https://doi.org/10.1007/s13349-023-00734-0.

[4] V. Périer, L. Dieng, L. Gaillet, and S. Fouvry, "Influence of an aqueous environment on the fretting behaviour of steel wires used in civil engineering cables," *Wear*, vol. 271, no. 9–10, pp. 1585–1593, Jul. 2011, doi: https://doi.org/10.1016/j.wear.2011.01.095.

[5] K. Feyrer, *Wire Ropes*. Springer Science & Business Media, 2007.

[6] L. Yang, Z. Liu, L. Ren, F. Liao, and M. Zuo, "Quantitative Detection of Local Flaw Under the Lift-Off Effect for Steel Wire Ropes," *IEEE Sensors Journal*, vol. 24, no. 16, pp. 26081–26090, Aug. 2024, doi: https://doi.org/10.1109/jsen.2024.3421650.

[7] P. Mazurek, "A Comprehensive Review of Steel Wire Rope Degradation Mechanisms and Recent Damage Detection Methods," *Sustainability*, vol. 15, no. 6, p. 5441, Mar. 2023, doi: https://doi.org/10.3390/su15065441.

[8] *Advances in Non-Destructive Testing Methods*. 2024. doi: https://doi.org/10.3390/books978-3-7258-1082-6.

[9] Y. Sun and Y. Kang, "Magnetic mechanisms of magnetic flux leakage nondestructive testing," *Applied Physics Letters*, vol. 103, no. 18, p. 184104, Oct. 2013, doi: https://doi.org/10.1063/1.4828556.

[10] J.-W. Kim and S. Park, "Magnetic Flux Leakage Sensing and Artificial Neural Network Pattern Recognition-Based Automated Damage Detection and Quantification for Wire Rope Non-Destructive Evaluation," *Sensors*, vol. 18, no. 2, p. 109, Jan. 2018, doi: https://doi.org/10.3390/s18010109.

[11] L. Ren, Z. Liu and J. Zhou, "Shaking Noise Elimination for Detecting Local Flaw in Steel Wire Ropes Based on Magnetic Flux Leakage Detection," *IEEE Transactions on Instrumentation and Measurement*, vol. 70, pp. 1-9, 2021, Art no. 3524909, doi: 10.1109/TIM.2021.3112792.

[12] S. Huang, Z. Wang, J. Yang, T. Gong, Z. Shan, and Y. Yang, "Adaptive fast Walsh-Hadamard transform for magnetic flux leakage signal of broken wire damage extraction under noise background," *Nondestructive Testing and Evaluation*, pp. 1–21, Mar. 2024, doi: https://doi.org/10.1080/10589759.2024.2325671.

[13] M. Wang, J. Li, and Y. Xue, "A New Defect Diagnosis Method for Wire Rope Based on CNN-Transformer and Transfer Learning," *Applied sciences*, vol. 13, no. 12, pp. 7069–7069, Jun. 2023, doi: https://doi.org/10.3390/app13127069.

[14] W. Yi, W. K. Chan, H. H. Lee, S. T. Boles, and X. Zhang, "An Uncertainty-Aware Deep Learning Model for Reliable Detection of Steel Wire Rope Defects," *IEEE Transactions on Reliability*, vol. 73, no. 2, pp. 1187–1201, Dec. 2023, doi: https://doi.org/10.1109/tr.2023.3335958.

[15] F. Pan, Y. Huang, L. Ren and Z. Liu, "Inspection of Wire Ropes Based on Magnetic Flux Leakage Images by Using YOLOv5, " *2023 Global Reliability and Prognostics and Health Management Conference (PHM-Hangzhou)*, Hangzhou, China, 2023, pp. 1-7, doi: 10.1109/PHM-Hangzhou58797.2023.10482526.

[16] S. Huang, L. Peng, H. Sun, and S. Li, "Deep Learning for Magnetic Flux Leakage Detection and Evaluation of Oil & Gas Pipelines: A Review," *Energies*, vol. 16, no. 3, p. 1372, Jan. 2023, doi: https://doi.org/10.3390/en16031372.

[17] F. Pan, Z. Liu, L. Ren, and M. Zuo, "Adaptive Local Flaw Detection Based on Magnetic Flux Leakage Images With a Noise Distortion Effect

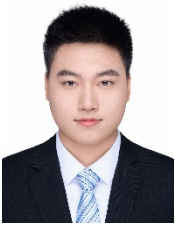

**Siyu You** (Student Member, IEEE) was born in Chengdu, Sichuan, China, in 2004. He is currently pursuing a B.S. degree in Electronic and Information Engineering at the University of Electronic Science and Technology of China (UESTC), Chengdu China.

His research interests include fault diagnosis, deep learning, computer vision, and radar sensing.

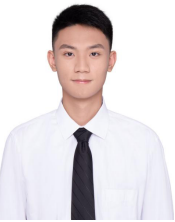

**Huayi Gou** was born in Chongqing, China, in 2004. He is currently pursuing a B.S. degree in electrical and electronic engineering with the University of Electronic Science and Technology of China, Chengdu, China.

His research interests include computer vision, non-destructive tests, signal processing and deep learning.

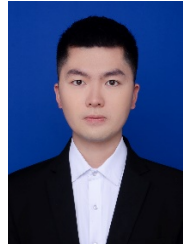

**Leilei Yang** (Graduate Student Member, IEEE) was born in Huaibei, Anhui, China. He received a master's degree in industrial engineering from the Chongqing University of Posts and Telecommunications, Chongqing, China, in 2021. He is currently pursuing a Ph.D. degree at the University of Electronic Science and Technology of China (UESTC), Chengdu, China. His research interests include magnetic detection sensor design and fault diagnosis.

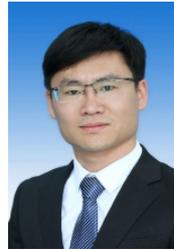

**Zhiliang Liu** (Senior Member, IEEE) received the Ph.D. degree from the School of Automation Engineering, UESTC, Chengdu, China, in 2013. From 2009 to 2011, he studied with the University of Alberta, Edmonton, AB, Canada, as a Visiting Scholar. From 2013 to 2015, he was an Assistant Professor at the School of Mechanical and Electrical Engineering, UESTC, where he has been an Associate Professor since 2015. His research interests include intelligent maintenance for complex machinery by using advanced signal processing and artificial intelligence methods.

Dr. Liu is a Fellow of the International Society of Engineering Asset Management (ISEAM), a global top 2% scientist, a candidate for the Sichuan Province Academic and Technical Leader, and a Senior Member of IEEE. He has published more than 120 academic papers, with six of them being recognized as highly cited papers by ESI, and two receiving Best Paper Awards. He has authored one academic monograph, applied for and authorized over 50 national invention patents, and transferred one patent technology for industrial use. Some of his research achievements have been applied to key engineering projects. Currently, Dr. Liu serves as an Associate Editor for IEEE Transactions on Instrumentation & Measurement and IEEE Sensors Journal, as well as a Youth Editorial Board member for Chinese flagship journals such as the Journal of Mechanical Engineering and the Journal of Southwest Jiaotong University.

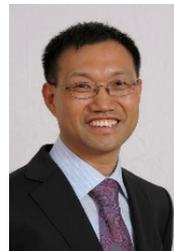

**Mingjian Zuo** (Fellow, IEEE) received the Bachelor of Science degree in Agricultural Engineering in 1982 from Shandong Institute of Technology, China; the Master of Science degree in 1986 and the Ph.D. degree in 1989 both in Industrial Engineering from Iowa State University, Ames, Iowa, U.S.A. He is Professor at Shandong University, Principal Scientist at Qingdao Mingserve Technology Ltd., and Professor Emeritus of the University of Alberta.

His research interests include system reliability analysis, maintenance modeling and optimization, signal processing, and fault diagnosis. He served as Department Editor of IISE Transactions; Associate Editor of IEEE Transactions on Reliability; Associate Editor of Journal of Risk and Reliability; Associate Editor of the International Journal of Quality, Reliability and Safety Engineering; Regional Editor of the International Journal of Strategic Engineering Asset Management, and was an Editorial Board member of Reliability Engineering and System Safety, Journal of Traffic and Transportation Engineering, and International Journal of Performability Engineering. He is a Fellow of the Canadian Academy of Engineering, a Fellow of the Institute of Industrial and Systems Engineers (IISE), a Fellow of the Engineering Institute of Canada (EIC), a Founding Fellow of the International Society of Engineering Asset Management (ISEAM), and Fellow of IEEE.